\documentclass[12pt,a4paper,fleqn]{article}
\usepackage[T1]{fontenc}
\usepackage{amsmath}
\usepackage{amssymb} % e.g. \gtrsim
\usepackage{bbm} % BlackBoeard letters
\usepackage[small]{caption2} % font of captions is different from text
\usepackage{graphicx} % \including PostScript
\usepackage{mathrsfs}	% e.g. nice Lagrange-L with \mathscr{L}
\usepackage[small,loose]{subfigure}  % subfigures with (a), (b) etc., ... and own caption
\usepackage{cite} % [1-5] instead of [1,2,3,4,5]	
\usepackage{cancel} 
\usepackage{color} 
\newcommand{\D}{\mathrm{d}}

\newcommand{\eVdist}{\kern-0.06em}
\newcommand{\mev}{\:\text{Me\eVdist V}}
\newcommand{\gev}{\:\text{Ge\eVdist V}}

\newcommand{\m}{\:\text{m}}
\newcommand{\cm}{\:\text{cm}}
\newcommand{\km}{\:\text{km}}
\newcommand{\fm}{\:\text{fm}}
\newcommand{\s}{\:\text{s}}
\newcommand{\yr}{\:\text{yr}}
\newcommand{\gyr}{\:\text{gyr}}

\DeclareMathAlphabet{\mathpzc}{OT1}{pzc}{m}{it}

\allowdisplaybreaks
\begin{document}
\thispagestyle{empty}
\begin{titlepage}

\begin{flushright}
TUM-HEP-804/11\\
MPP-2011-37
\end{flushright}

\vspace*{1.0cm}

\begin{center}
\Huge\textbf{New Limits on Dark Matter from Super-Kamiokande}
\end{center}
\vspace{1cm}
 \center{
\textbf{
Rolf Kappl\footnote[1]{Email: \texttt{rolf.kappl@ph.tum.de}}$^{,ab}${}, 
Martin Wolfgang Winkler\footnote[2]{Email: \texttt{mwinkler@ph.tum.de}}$^{,a}$
}
}
\\[5mm]
\begin{center}
\textit{$^a$\small
~Physik-Department T30, Technische Universit\"at M\"unchen, \\
James-Franck-Stra\ss e, 85748 Garching, Germany
}
\\[5mm]
\textit{$^b$\small
~Max-Planck-Institut f\"ur Physik (Werner-Heisenberg-Institut),\\
F\"ohringer Ring 6,
80805 M\"unchen, Germany
}
\end{center}

\date{}
\vspace{1cm}

\begin{abstract}
The signals observed at the direct detection experiments DAMA, CoGeNT and CRESST could be explained by light WIMPs with sizeable spin-independent cross sections with nucleons. The capture and subsequent annihilation of such particles in the sun would induce neutrino signals in the GeV range which may be observed at Super-Kamiokande. We determine the rate of upward stopping muons and fully contained events at Super-Kamiokande for various possible WIMP annihilation channels. This allows us to provide strong constraints on the cross section of WIMPs with nucleons. We find that the DAMA and CoGeNT signals are inconsistent with standard thermal WIMPs annihilating dominantly into neutrino or tau pairs. We also provide limits for spin-dependent WIMP nucleus scattering for masses up to 80 GeV. These exclude the DAMA favored region if WIMPs annihilate even subdominantly into neutrinos, taus, bottoms or charms.
\end{abstract}

\end{titlepage}

\newpage

\section{Introduction}

Throughout the last decades overwhelming evidence for the existence of dark matter has been collected. Despite the precise understanding of the gravitational effects of dark matter, its nature still remains a secret.
Among the most promising dark matter candidates is a particle with weak scale interactions and mass (WIMP). Its abundance from thermal production in the early universe would naturally match the dark matter abundance. 

For several years great efforts were made to detect WIMPs directly by their interaction with nuclei. Recently, the three direct detection experiments DAMA~\cite{Bernabei:2008yi,Bernabei:2010mq}, CoGeNT~\cite{Aalseth:2010vx} and CRESST~\cite{Seidel:2010nn} have announced signals which can consistently be explained by elastic scatterings of rather light WIMPs with masses of a few $\mathrm{GeV}$~\cite{Hooper:2010uy}. This interpretation is now challenged by a low-threshold analysis from the CDMS collaboration which excludes the DAMA and a large fraction of the CoGeNT favored region in parameter space~\cite{Ahmed:2010wy}. Nevertheless, one should be cautious to discard the light dark matter hypothesis as several concerns about the CDMS analysis have been raised~\cite{Collar:2011kf}. Furthermore, even if the CDMS limit persists light dark matter may still explain at least the CoGeNT and CRESST results.\footnote{A precise determination of the parameter space where the CRESST excess can be explained by dark matter would require further knowledge especially of the neutron and alpha backgrounds.}

In any case it seems reasonable to consider independent measures to search for light dark matter in collider- and astrophysics. In the past, analysis of the Tevatron and LEP data lead to constraints on possible WIMP couplings to quarks, gluons and electrons~\cite{Goodman:2010yf,Goodman:2010ku,Fox:2011fx}. Observations of the diffuse gamma ray spectrum~\cite{Abazajian:2010sq,Abdo:2010dk,Hutsi:2010ai}, the antiproton spectrum~\cite{Lavalle:2010yw} and the cosmic microwave background~\cite{Galli:2009zc} were used to put limits on the annihilation cross section of light WIMPs. Most remarkably, the latest data from WMAP seem to disfavor a standard thermal WIMP in the GeV range unless it annihilates predominantly into muons, taus or neutrinos, or its annihilation cross section is velocity suppressed~\cite{Hutsi:2011vx}. Neutrino telescopes offer a complementary probe of the light dark matter hypothesis. They can detect the neutrino signals from dark matter annihilation and are especially sensitive to the neutrino and tau channels.

In this study we consider WIMP capture and annihilation in the sun including also a treatment of velocity suppressed annihilation. We use the observed neutrino spectrum at Super-Kamiokande to constrain the WIMP nucleon cross section for various possible WIMP annihilation channels. While the Super-Kamiokande collaboration analyzed upward through-going muons in the dark matter context~\cite{Desai:2004pq}, we focus on fully contained events and upward stopping muons. These event categories are known to be more relevant for light WIMPs~\cite{Feng:2008qn,Kumar:2009af,Andreas:2009hj} but have not been used to constrain WIMP cross sections in a model-independent way so far.\footnote{In~\cite{Niro:2009mw} upward stopping muons have been used to constrain specific MSSM neutralinos.} (We also note that an earlier analysis on light WIMPs~\cite{Hooper:2008cf} which is also used in~\cite{Fitzpatrick:2010em} contains a serious error and should be taken with care.\footnote{With their formula~14 the authors of~\cite{Hooper:2008cf} calculate the WIMP induced rate of upward going muons in Super-Kamiokande. But to derive their limits, they compare it with Super-Kamiokande data on upward through-going muons. Thereby, they overlook that upward going muons do not necessarily traverse the entire detector volume. Indeed, in the energy range considered most observed muons are created and/or stopped inside the detector (see figure~1 in~\cite{Ashie:2005ik}).})

We obtain strong limits on the spin-independent WIMP nucleon cross section which we relate to direct detection experiments. We find that -- for standard astrophysical assumptions -- a thermal WIMP which annihilates dominantly into tau or neutrino pairs is excluded as a possible source of the DAMA and CoGeNT excesses unless the cross section is velocity suppressed. But even for pure p-wave annihilation at least the DAMA region is ruled out for the two channels. We will comment on models which can account for the DAMA and CoGeNT signals and are compatible with the combined data from indirect dark matter detection.

We also provide constraints on the spin-dependent WIMP proton cross section for masses up to $80\gev$. Compared to direct detection experiments they are more stringent for all considered annihilation channels.

\section{Theoretical Framework}

In this study we constrain the WIMP nucleon cross section by considering WIMP induced neutrino signals from the sun. The neutrino spectrum from dark matter annihilation depends strongly on the underlying particle physics model. We focus on light WIMPs $m_\chi \leq 80\gev$ but otherwise keep our analysis as general as possible. Therefore, we derive our limits independently for various possible WIMP annihilation channels. 

\subsection{Annihilation Channels}

We consider the annihilation of WIMPs into pairs of Standard Model fermions
\begin{equation}\label{eq:SMchannels}
 \chi\chi \rightarrow b\bar{b}, \,c\bar{c},\,\tau\bar{\tau},\, \nu\bar{\nu}\;.
\end{equation}
For the neutrino channel we assume $\nu = (\nu_\tau +\nu_\mu + \nu_e)/3$, but -- due to neutrino oscillations -- other combinations would lead to similar limits. Annihilation into photons, lighter quarks, gluons, electrons and muons is omitted as the corresponding neutrino signals would be negligible.\footnote{Annihilation into muons would of course induce a large number of neutrinos. Their energy would, however, reside below the detection threshold as muons in the sun -- due to their long life-time -- are stopped before they decay.}
We further take into account
\begin{equation}\label{eq:newchannel}
 \chi\chi \rightarrow 4\tau\;,
\end{equation} 
where we assume that the energy is distributed equally between the taus. This channel is e.g. accessed if $\chi\chi$ annihilates into a pair of new gauge or Higgs bosons which subsequently decay into taus. We include it as neutrino telescopes are very sensitive to WIMPs annihilating into taus and we want to determine whether constraints can be alleviated if the energy is shared between more taus.
The neutrino spectra resulting from~\eqref{eq:SMchannels} and~\eqref{eq:newchannel} were calculated with PYTHIA Monte Carlo (version 8.1)~\cite{Sjostrand:2007gs}. We checked that they agree well with those given in~\cite{Cirelli:2005gh,Blennow:2007tw}, only for the tau neutrino production in the \(c\bar{c}\) channel there is a discrepancy. This can, however, easily be explained as the spectra of~\cite{Cirelli:2005gh,Blennow:2007tw} were calculated with an older PYTHIA version~\cite{Sjostrand:2006za} which uses an out-dated value for the semileptonic branching ratio of \(D_s^{\pm}\) mesons.

\subsection{Annihilation Cross Section}

Throughout this study we assume that dark matter is produced thermally. In this case the annihilation cross section $\sigma v_\text{rel}$ can be related to the dark matter abundance. Expanding $\sigma v_\text{rel}$ in the relative WIMP velocity $\sigma v_\text{rel} = a + b v_\text{rel}^2 +\mathcal{O}(v_\text{rel}^4)$ one finds~\cite{Drees:2009bi}\footnote{This formula is not applicable to models with resonant annihilation, coannihilations or Sommerfeld enhancement.}
\begin{equation}\label{eq:Omega}
 \Omega_\chi \,h^2 
 = 1.0\times 10^{-27}\cm^3\s^{-1}\,
 \frac{m_\chi}{\sqrt{g_*(T_F)}\,T_F\,(a + 3\, T_F\,b/m_\chi)}\;,
\end{equation}
where $g_*$ denotes the effective number of relativistic degrees of freedom, $T_F$ the freeze-out temperature and \(m_\chi\) the WIMP mass. We set $T_F = m_\chi / 20$ which is a reasonable approximation. 
In case of s-wave annihilation $\sigma v_\text{rel}$ is determined by the first term in the velocity expansion, in case of p-wave annihilation by the second term. The thermal WIMP abundance matches the dark matter abundance for
\begin{subequations}
 \begin{align}
  a&= \frac{1.8 \cdot 10^{-25}\cm^3\s^{-1}}{\sqrt{g_*(\frac{m_\chi}{20})}} \qquad \text{(s-wave annihilation)}\;,\\
  b&= \frac{1.2 \cdot 10^{-24}\cm^3\s^{-1}}{\sqrt{g_*(\frac{m_\chi}{20})}} \qquad \text{(p-wave annihilation)}\;,
 \end{align}
\end{subequations}
where we take $g_*$ from~\cite{Gondolo:1990dk}. We will discuss s- and p-wave annihilation separately in the next section.

\section{Dark Matter in the Sun}\label{sec:darkmattersun}

WIMPs on their way through the sun may get gravitationally bound if they lose energy by scattering with nuclei. The trapped WIMPs can annihilate or escape via evaporation. A coherent treatment of the three processes which is valid at low WIMP masses and includes velocity suppressed annihilation is rare in the literature. Therefore, we find it appropriate to discuss them in some detail. Throughout this study we will use the solar model AGSS09~\cite{Serenelli:2009yc} for the composition and properties of the sun.

\subsection{Capture}
A WIMP which scatters off a nucleus at radius $r$ inside the sun gets trapped if its velocity after the scattering is lower than the escape velocity $v_\text{esc}(r)$. Let us first discuss spin-independent WIMP nucleus scattering. For this the differential contribution of nuclei $i$ at radius $r$ to the capture rate takes the form~\cite{Wikstrom:2009kw}
\begin{equation}\label{eq:capture}
 \frac{dC_{\odot,i}}{dV} = \frac{\rho_\chi \,\rho_{\odot,i}(r) }{2 m_\chi \mu_i^2} \,\sigma_i\, \int\limits_{0}^{\infty} du\:\frac{f(u)}{u} \int\limits_{E_{R,\text{min}}}^{E_{R,\text{max}}} dE_R\:|F(E_R)|^2\;.
\end{equation}
Here $\rho_{\odot,i}(r)$ is the mass density of nuclei $i$ at radius $r$ in the sun and $\mu_i$ the reduced mass of the WIMP nucleus system. The WIMP nucleus cross section $\sigma_i$ is defined as
\begin{equation}
\sigma_i = \sigma_p \,A_i^2\, \frac{\mu_i^2}{\mu_p^2}\;,
\end{equation}
where $A_i$ is the mass number of the nucleus $i$, $\sigma_p$ and $\mu_p$ the WIMP proton cross section and reduced mass. We assumed equal couplings to protons and neutrons.
From astrophysics it enters the local dark matter density $\rho_\chi \simeq 0.3 \gev\cm^{-3}$ and the velocity distribution $f(u)$ of WIMPs outside the gravitational potential of the sun. We use a Maxwellian distribution\footnote{We neglect that the velocity distribution should be truncated at the local galactic escape velocity as this has virtually no impact on capture rates.} shifted by the velocity of the sun $v_\odot$:
\begin{equation}
 \frac{f(u)}{u}=\frac{1}{\sqrt{\pi}\,v_\odot^2}\left( e^{-(u-v_\odot)^2/v_\odot^2}-e^{-(u+v_\odot)^2/v_\odot^2}\right)\;
\end{equation}
with $v_\odot=220\km\s^{-1}$. Note that the velocity of an incoming WIMP (before scattering) inside the sun is simply given by $\sqrt{u^2+v_\text{esc}^2(r)}$. 

Finally, decoherence effects in the scattering enter via the form factor $F(E_R)$, where $E_R$ is the recoil energy of the nucleus. While $F(E_R)=1$ for all relevant energies in the case of hydrogen, the suppression can be substantial for heavier nuclei like iron. We use Gaussian form factors~\cite{Gould:1987ir}
\begin{equation}
 |F(E_R)|^2 = e^{-E_R/E_i}
\end{equation}
with
\begin{align}
 E_i &= \frac{3}{2 m_i R_i^2}\;,\\
 R_i &= (0.9\,A^{1/3} + 0.3)\fm\;,
\end{align}
where $m_i$ is the mass and $R_i$ the RMS charge radius of the nucleus $i$. It is known that Gaussian form factors are less accurate than e.g. Woods-Saxon form factors especially at large $E_R$. They have, however, the advantage that they can be integrated analytically. Further, as only the integrated form factors enter~\eqref{eq:capture}, the error we make is at the percent level~\cite{Ellis:1991ef}. The $E_R$ integration runs from the minimal energy transfer $E_{R,\text{min}}$ required to trap the WIMP to the maximal energy transfer $E_{R,\text{max}}$ allowed by kinematics. One finds
\begin{equation}
 E_{R,\text{min}}= \frac{1}{2} m_\chi u^2\;,\qquad
 E_{R,\text{max}}=\frac{2 \mu_i^2 }{m_i}\,(u^2+v_\text{esc}^2(r))\;.
\end{equation}
The total capture rate is given by
\begin{equation}
 C_\odot = 4\pi\sum\limits_i \int\limits_0^{R_\odot} dr \:r^2\: \frac{dC_{\odot,i}}{dV}\;,
\end{equation}
where $R_\odot$ is the solar radius. The sum runs over all types of nuclei in the sun. We consider those from hydrogen up to nickel.

In the case of spin-dependent scattering the WIMP nucleus cross section does not experience the coherent enhancement $\propto A^2$. Therefore only capture at hydrogen is relevant. Its differential contribution to capture rate is given by~\eqref{eq:capture} with the form factor $F(E_R)$ replaced by $1$.

\begin{figure}[h!]
\centering
  \includegraphics[width=8.5cm]{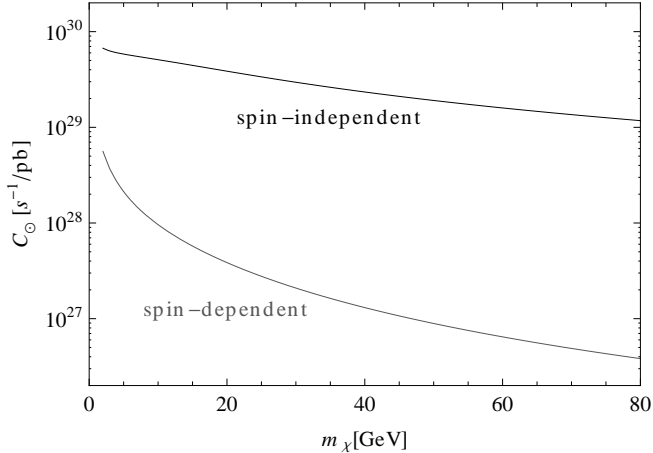}
\vspace{-2mm}
\caption{Solar capture rate for spin-independent and spin-dependent WIMP nucleus scattering.}
\label{fig:capture}
\end{figure}

The capture rates for spin-independent and spin-dependent scattering are shown in figure~\ref{fig:capture}.

\subsection{Annihilation}\label{sec:annihilation}

WIMPs which are trapped in the sun thermalize and, therefore, follow a Maxwellian velocity distribution with temperature $T_\chi$:
\begin{equation}\label{eq:thermaldistribution}
 f(v) = \sqrt{\frac{2 m_\chi^3}{\pi T_\chi^3}} v^2 e^{-(m_\chi v^2)/(2T_\chi)}\;.
\end{equation}
It turns out that for all relevant masses the WIMPs reside close to the center of the sun. In this region the variations of the solar temperature $T_\odot(r)$ and density $\rho_\odot(r)$ are small and one can approximate both quantities by constants. A reasonable choice is $T_\chi=T_\odot(\bar{r})$ and $\rho_\odot(r)=\rho_\odot(\bar{r})$ where $\bar{r}$ is the mean WIMP orbit radius. To derive $\bar{r}$ we use that the WIMP number density scales as~\cite{Griest:1986yu}
\begin{equation}\label{eq:wimpdensity}
 n_\chi(r)=n_0 e^{-(m_\chi \phi(r))/T_\chi}\,.
\end{equation}
For constant density, the gravitational potential with respect to the solar core reads $\phi(r)=2\pi\rho_{\odot}(\bar{r}) r^2 G/3$ with $G$ being Newton's constant. By averaging $r$ over the WIMP density we arrive at the implicit equation
\begin{equation}
 \bar{r} = \sqrt{\frac{6\, T_\odot(\bar{r})}{\pi^2 G \rho_{\odot}(\bar{r})\, m_\chi}}
\end{equation}
which can be solved numerically.

If we denote the total WIMP number by $N$, their annihilation rate is given by $\Gamma_\odot=A_\odot N^2 /2$ with
\begin{equation}
 A_\odot = \frac{1}{N^2}\; \int dr\; 4\pi r^2\, n_\chi^2(r) \,\langle \sigma v_\text{rel} \rangle _\odot = \left(\frac{\sqrt{2}}{\pi \bar{r}} \right)^3\langle \sigma v_\text{rel} \rangle _\odot\;.
\end{equation}
We find that for WIMP masses $m_\chi \gtrsim 1\gev$ the rate is well approximated by
\begin{equation}
 A_\odot = 4.5 \cdot 10^{-30}\cm^{-3} \left(\frac{m_\chi-0.6\gev}{10\gev}\right)^{3/2}\langle \sigma v_\text{rel} \rangle_\odot\;.
\end{equation}
If we perform the thermal averaging for an annihilation cross section of the form $\sigma v_\text{rel} = a + b\, v_\text{rel}^2$ we find
 \begin{equation}
 \langle \sigma v_\text{rel} \rangle_\odot = a + \frac{6\,T_{\odot}(\bar{r})}{m_\chi}\,b\,.
\end{equation}

\subsection{Evaporation}

Trapped WIMPs may escape from the sun if they gain energy by scattering with nuclei inside the sun~\cite{Griest:1986yu,Gould:1987ju,Gould:1989tu}. This process -- called evaporation -- is highly sensitive to the WIMP mass. The evaporation rate per WIMP roughly scales as
\begin{equation}
 E_\odot \sim \frac{1}{t_\odot} e^{ -30\,(m_\chi-m_\text{evap})/m_\text{evap} }\;,
\end{equation}
where $t_\odot\simeq 4.7 \gyr$ is the age of the sun and $m_\text{evap}$ the evaporation mass. The latter is defined as the mass for which $E_\odot=1/t_\odot$. Note that if $m_\chi$ exceeds $m_\text{evap}$ by a few percent, evaporation is totally negligible. Conversely, if $m_\chi$ falls slightly below $m_\text{evap}$ virtually all trapped WIMPs escape the sun via evaporation. The evaporation rate can be estimated as~\cite{Gould:1989tu}
\begin{equation}\label{eq:evaporation}
 E_\odot \simeq \frac{8}{\pi^3} \frac{\sigma_\text{evap}}{\bar{r}^3}\,\bar{v}\,\frac{E_\text{esc}}{T_{\odot}(\bar{r})}\; \exp{\left[ -\frac{E_\text{esc}}{2T_{\odot}(\bar{r})}\right] }\;.
\end{equation}
Here $E_\text{esc}$ denotes the escape energy at the center of the sun and $\sigma_\text{evap}$ the evaporation cross section. The latter is given by the total cross section of all nuclei interior to the radius $r_{0.95}$ which is defined by $T_{\odot}(r_{0.95})=0.95\,T_{\odot}(\bar{r})$. Finally the mean WIMP speed $\bar{v}$ for the thermal distribution~\eqref{eq:thermaldistribution} reads
\begin{equation}
 \bar{v}=\sqrt{\frac{8 T_{\odot}(\bar{r})}{\pi m_\chi}}\;.
\end{equation}
Note that the estimate~\eqref{eq:evaporation} is only accurate to within a factor of three~\cite{Gould:1989tu}.\footnote{The error comes from the fact that the WIMP velocity distribution in the sun shows deviations from a thermal distribution at high velocities which is not taken into account in~\eqref{eq:evaporation}. See discussion in~\cite{Gould:1987ju}.} This precision is, however, sufficient for our purposes as it translates into an uncertainty in the evaporation mass of only $3\%$.
We find that the evaporation mass is well approximated by
\begin{equation}
 m_\text{evap} = m_0 + 0.32\gev \:\log_{10}\left(\frac{\sigma_p}{10^{-40}\cm^2} \right)\;,
\end{equation}
where $m_0=3.5\gev$ ($m_0=3.02\gev$) in the case of spin-independent (spin-dependent) interactions.

\subsection{Total WIMP Number and Annihilation Signal}

The evolution of the total WIMP number in the sun is described by the differential equation
\begin{equation}\label{eq:numberevolution}
 \dot{N}=C_\odot - A_\odot N^2 - E_\odot N
\end{equation}
which takes into account capture, annihilation and evaporation. The general solution to this equation is rather unhandy, but in case evaporation can be neglected it takes the simple form
\begin{equation}
 N=\sqrt{\frac{C_\odot}{A_\odot}} \tanh(\sqrt{C_\odot A_\odot}\,t)\,
\end{equation}
and the total annihilation rate at present times reads
\begin{equation}
 \Gamma_\odot=\frac{1}{2} A_\odot N^2 = \frac{1}{2} C_\odot \tanh^2{\left(\sqrt{C_\odot A_\odot} \:t_\odot \right) }\;.
\end{equation}
In the limit $\sqrt{C_\odot A_\odot} \:t_\odot \gg 1$ equilibrium between capture and annihilation is reached and one finds $\Gamma_\odot = C_\odot/2$. For a capture rate of $10^{25}\s^{-1}$ this requires $\langle \sigma v_\text{rel} \rangle_\odot\gg 10^{-30} \cm^3\s^{-1}$. Usually, it is assumed that the leading contribution to the annihilation cross section is velocity-independent, i.e. $\sigma v_\text{rel} = a$ with $a=\mathcal{O}(10^{-26}\cm^3\s^{-1})$ (s-wave annihilation). In this case equilibrium is safely reached. 

A viable exception is the case where $\sigma v_\text{rel}$ at freeze-out is dominated by the second term in the velocity expansion, i.e. $\sigma v_\text{rel} = a + b \,v_\text{rel}^2$ with $a\ll b$. Then, one has to take into account that the velocity distribution of WIMPs in the sun is different from the distribution at freeze-out. For pure p-wave annihilation ($a=0$) the thermally averaged annihilation cross section in the sun is reduced by a factor $T_\odot(\bar{r})/T_F$ compared to freeze-out. In this case capture annihilation equilibrium is typically not reached and $\Gamma_\odot < C_\odot/2$. We decided to derive our limits separately for s- and p-wave annihilation. 

Note, however, that even if $a\ll b$, the small $a$ may still be sufficient to put capture and annihilation into equilibrium in the sun. If this is the case the limits from s-wave annihilation apply (with some modifications very close to the evaporation mass).

\section{Neutrino Propagation and Interactions}

Neutrino telescopes like Super-Kamiokande are able to detect muon neutrinos from all directions of the sky. In order to determine the incoming muon neutrino signal from dark matter annihilation one has to consider neutrino oscillations and interactions.

\subsection{Neutrino Propagation}

The muon neutrino flux which arrives at the earth from WIMP annihilation in the sun may be written as
\begin{equation}
\frac{\D F_{\nu}}{\D E_{\nu}}=
\frac{\Gamma_{\odot}}{4\pi
  d_\odot^2}\left.\frac{\D N_{\nu_\mu}}{\D E_{\nu_\mu}}\right|_{
\text{earth}}\;,
\end{equation}
where $d_\odot$ denotes the distance between earth and sun, $(\D N_{\nu_\mu}/\D E_{\nu_\mu})|_\text{earth}$ the propagated muon neutrino spectrum per annihilation. We use the density matrix formalism as described in~\cite{Raffelt:1992uj,Cirelli:2005gh,Strumia:2006db} to derive $(\D N_{\nu_\mu}/\D E_{\nu_\mu})|_\text{earth}$ from the initial neutrino spectra in the sun $(\D N_{\nu_{e,\mu,\tau}}/\D E_{\nu_{e,\mu,\tau}})|_\odot$.

The evolution equation used for the propagation from the production point in the solar core to the earth reads
\begin{equation}\label{eq:evolution}
\frac{\D\varrho}{\D r}=-i[H,\varrho]+
\left.\frac{\D \varrho}{\D r}\right|_{\text{NC}}+
\left.\frac{\D \varrho}{\D r}\right|_{\text{CC}}-\epsilon[H,[H,\varrho]]\;.
\end{equation} 
The diagonal entries of the density matrix $\varrho$ can be identified with the three neutrino flavor eigenstates while off-diagonal entries correspond to superpositions of flavors. The terms on the right hand side describe neutrino oscillations, neutral and charged current interactions and decoherence effects respectively. For their definition see~\cite{Cirelli:2005gh,Strumia:2006db}.\footnote{We neglect \(\nu_{\tau}\) regeneration in $(\D \varrho/\D r)|_{\text{CC}}$ as it is subleading for low energetic neutrinos~\cite{Niro:2009mw}.} The evolution equation has to be solved numerically, the density matrix at $r=0$ is given by the initial neutrino spectra obtained from PYTHIA. Neutrino interactions will be discussed in more detail in the next section.

\subsection{Neutrino Interactions}\label{sec:neutrinointeractions}

Neutrino charged and neutral current interactions with nuclei at high energies ($E_\nu \gtrsim 10\gev$) are dominated by deep inelastic scattering. In this regime simple analytic approximations for the neutrino cross sections exist (see e.g. Appendix~C in~\cite{Barger:2007xf}). These were continuously used in earlier analyses which mostly dealt with neutrino signals from WIMPs of several hundred GeV. 

In this study, however, we consider the annihilation of light WIMPs ($m_\chi \leq 80\gev$) which typically induces neutrinos with energies of a few GeV or even below. For low energy interactions, the momentum transfer between neutrinos and nuclei is so small that the deep inelastic prescription badly fails. In fact, the neutrino cross section at energies $E_\nu \lesssim 3\gev$ is dominated by quasi-elastic scattering~\cite{LlewellynSmith:1971zm} and single pion production which proceeds through an intermediate nucleon resonance~\cite{Fogli:1979cz,Rein:1980wg}. Especially the latter makes an analytic treatment of neutrino interactions difficult as a series of resonances have to be considered. We therefore decided to simulate neutrino interactions with the publically available neutrino event generator NEUGEN 3.5.5~\cite{Gallagher:2002sf} which automatically takes into account the discussed processes. A solid treatment of low energy interactions is especially important for the calculation of neutrino detection rates, but we also take the neutrino cross sections which enter the propagation equation~\eqref{eq:evolution} from NEUGEN.

\section{Neutrino Detection at Super-Kamiokande}

Super-Kamiokande is the most sensitive experiment for neutrino signals in the GeV range as they arise from light dark matter annihilation. Therefore, we will not consider other neutrino telescopes. Super-Kamiokande is a water Cherenkov detector of cylindrical shape (height: $36.2\m$, radius: $16.9\m$). It is located in the Kamioka-mine ($36^\circ\:14^\prime\:\text{N}$, $137^\circ\:11^\prime\:\text{E}$) and surrounded by rock. 
The detection rates at Super-Kamiokande depend on the energy and the zenith angle\footnote{Meant is the angle between neutrino and cylinder axis of Super-Kamiokande.} $\theta_\nu$ of the incoming neutrinos. The double differential muon neutrino flux from solar dark matter annihilation can be written as
\begin{equation}
 \frac{\D F_{\nu}}{\D E_{\nu}\D\cos\theta_{\nu}} = \frac{\D F_{\nu}}{\D E_{\nu}} \:\frac{\D P}{\D\cos\theta_{\nu}}\;.
\end{equation}
The function $\D P/\D\cos\theta_{\nu}$ denotes the differential probability for the neutrino beam to arrive with angle $\cos\theta_{\nu}$ (averaged over the run time of Super-Kamiokande). It can be obtained by tracking the orientation of the Super-Kamiokande detector relative to the sun taking into account the orbital and rotational movement of the earth. We employed the IDL Astronomy User's Library~\cite{Landsman:1993aa} to calculate $\D P/\D\cos\theta_{\nu}$, the run time of Super-Kamiokande was taken from figure~4.16 in~\cite{Ishihara:2010zzb}.

\begin{figure}[h!]
\centering
  \includegraphics[width=8cm]{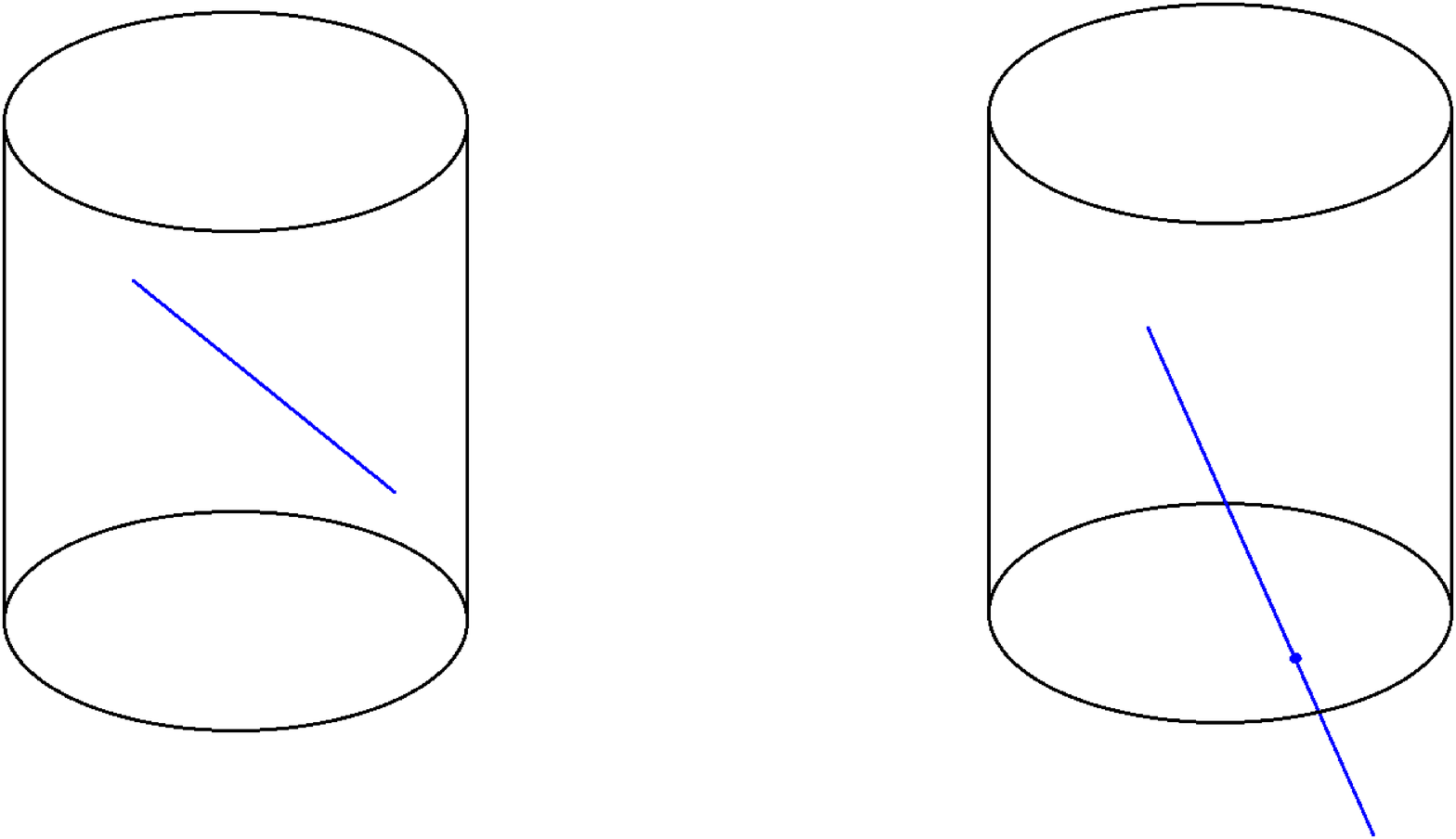}
\vspace{-4mm}
\caption{Illustration of event types at Super-Kamiokande: fully contained events (left), upward stopping muons (right).}
\label{fig:eventtypes}
\end{figure}

Super-Kamiokande is searching for neutrino induced muons in several event categories. We will focus on fully contained muon-like events and upward stopping muons (see figure~\ref{fig:eventtypes}).

\subsection{Fully Contained Events}

An event is categorized as fully contained muon-like if a particle identified as muon is created and stopped within the detector. To eliminate backgrounds, several cuts are applied by the Super-Kamiokande collaboration. Events are discarded if the initial vertex is less than $2\m$ away from the detector wall or if the momentum of the muon is below a threshold momentum $p_{\mu,\text{th}}$. The latter is $p_{\mu,\text{th}}=200\mev$ for single-ring and $p_{\mu,\text{th}}=600\mev$ for multi-ring events~\cite{Ashie:2005ik} which corresponds to threshold energies of $E_{\mu,\text{th}}=226\mev$ and $E_{\mu,\text{th}}=609\mev$ respectively. Here single-ring stands for events with one observed Cherenkov ring in the detector, multi-ring for those with at least two Cherenkov rings. Multi-ring events are induced if the neutrino interaction gives rise to further charged particles (pions) apart from the muon. 

We identify quasi-elastic interactions as single-ring all other types of charged current interactions as multi-ring. By this we neglect the finite probability that events involving pions are ``misidentified'' as single-ring.  As multi-ring events have a stronger momentum cut we thus discard slightly too many events. This, however, only makes our limits more conservative.

The fully contained muon-like event rate can be written as
\begin{equation}
\begin{split}
\label{eq:FCrate}
 R^\text{FC} &= n_{\text{H$_2$O}} \int\limits_0^{m_\chi}\! \D E_\nu \int\limits_{-1}^1\! \D\cos\theta_\nu \:\frac{\D F_\nu}{\D E_\nu\: \D\cos\theta_\nu} \,\int\limits_{\text{V}_{\text{-2m}}} \! \D^3x \int\limits_{E_{\mu,\text{th}}}^{E_\nu} \! \D E_\mu
\\
&\quad\times\, \int\limits_{-1}^1\! \D\cos\theta_{\mu\nu}\:\frac{1}{2\pi} \int\limits_0^{2\pi} \! \D\varphi_{\mu\nu} \:\frac{\D\sigma_{\text{cc}}^{\text{H$_2$O}}}{\D E_\mu \:\D\cos\theta_{\mu\nu}}\;\delta^{\text{FC}}\left(\mathbf{x},E_\mu,\cos\theta_\mu(\theta_{\mu\nu},\varphi_{\mu\nu})\right)\;.
\end{split}
\end{equation}
Here $n_{\text{H$_2$O}}$ is the number density of water molecules and $\mathbf{x}$ the position of the interaction vertex. The space integral runs over all points inside the detector volume which are more than $2\m$ away from the detector wall. The function $\delta^{\text{FC}}=1$ if the muon stops inside the detector, otherwise $\delta^{\text{FC}}=0$. Taking the track length of muons in water from table II-28 in~\cite{Groom:2001kq}, $\delta^{\text{FC}}$ can be determined by simple geometrical considerations.
We further denote the zenith angle of the incoming neutrino and outgoing muon by $\theta_\nu$ and $\theta_\mu$ respectively. The angles $\theta_{\mu\nu},\varphi_{\mu\nu}$ are the zenith and azimuth angles of the muon relative to the neutrino, i.e.
\begin{equation}
 \cos\theta_\mu(\theta_{\mu\nu},\varphi_{\mu\nu}) = \cos\theta_\nu  \,\cos\theta_{\mu\nu} +  \sin\varphi_{\mu\nu} \,\sin\theta_\nu\,  \sin\theta_{\mu\nu}\;.
\end{equation}
If $\theta_{\mu\nu}=\varphi_{\mu\nu}=0$ this would imply that neutrino and muon are collinear.

The differential neutrino charged current cross section at water molecules $\D\sigma_{\text{cc}}^{\text{H$_2$O}}$ is calculated with NEUGEN (see section~\ref{sec:neutrinointeractions}). Note that in most analyses performed so far, the approximation $\theta_\mu = \theta_\nu$ was used. This would simplify~\eqref{eq:FCrate} substantially, but -- unfortunately -- the approximation is not valid for the energy range we consider. Therefore, we keep track of the finite angle between neutrino and muon.

As a cross-check for our approach we used the HKKMS fluxes~\cite{Honda:2006qj} to calculate the (non-oscillated) prediction for fully contained events from atmospheric neutrinos in Super-Kamiokande~I. We found good agreement with~\cite{Ashie:2005ik} concerning the zenith angle distribution of events. We systematically underestimate the number of events by a few percent. This is what we expect by our implementation of the event cuts (see above).

\subsection{Upward stopping muons}

Super-Kamiokande can also detect muons which come from neutrino interactions in the rock surrounding the detector. To separate neutrino induced muons from cosmic ray muons only those traveling in the upward direction are considered. Super-Kamiokande distinguishes between muons which traverse the entire detector volume (upward through-going muons) and muons which stop inside the detector (upward stopping muons). For both the energy threshold is $E_{\mu,\text{th}}=1.6\gev$, but the latter are more important for constraining light dark matter and we focus on these. Their event rate is given by
\begin{equation}
\begin{split}
\label{eq:ratestopping}
 R^\text{Stop} &= \frac{\rho_{\text{Rock}}}{m_\text{Si}} \int\limits_0^{m_\chi}\! \D E_\nu \int\limits_{-1}^1\! \D\cos\theta_\nu \:\frac{\D F_\nu}{\D E_\nu\: \D\cos\theta_\nu} \,\int\limits_{\text{V}_{\text{Rock}}} \! \D^3x \int\limits_{E_{\mu,\text{th}}}^{E_\nu} \! \D E_\mu
\\
&\quad\times\, \int\limits_{-1}^1\! \D\cos\theta_{\mu\nu}\:\frac{1}{2\pi} \int\limits_0^{2\pi} \! \D\varphi_{\mu\nu} \:\frac{\D\sigma_{\text{cc}}^{\text{Si}}}{\D E_\mu \:\D\cos\theta_{\mu\nu}}\;\delta^{\text{Stop}}\left(\mathbf{x},E_\mu,\cos\theta_\mu(\theta_{\mu\nu},\varphi_{\mu\nu})\right)\\
&\quad\times\: \Theta(-\cos\theta_\mu(\theta_{\mu\nu},\varphi_{\mu\nu}))\;.
\end{split}
\end{equation}
The mass density of rock $\rho_{\text{Rock}}$ can be taken from table IV-6 of~\cite{Groom:2001kq}. In the same reference one can find the energy loss of muons in rock. For simplicity we have assumed that rock consists entirely of silicon\footnote{The actual composition of rock is hardly relevant for interaction rates. Standard rock is known to consist to equal parts of protons and neutrons~\cite{Groom:2001kq}. All nuclei $A_i$ with an equal number of protons and neutrons have virtually the same ratio $\sigma_{\text{cc}}^{A_i}/m_{A_i}$.}, $m_\text{Si}$ denotes the mass of a silicon nucleus and $\D\sigma_{\text{cc}}^{\text{Si}}$ the differential neutrino charged current cross section at silicon. 
The space integration runs over the rock surrounding the detector. Again, the function $\delta^{\text{Stop}}=1$ if the muon stops inside the detector and $\delta^{\text{FC}}=0$ otherwise. The Heaviside function in the last line of~\eqref{eq:ratestopping} takes care that downward going muons are discarded.

Again we used the HKKMS fluxes to verify that we can reproduce the Monte Carlo prediction for upward stopping muons  in Super-Kamiokande~I with good precision.

\section{Limits from Super-Kamiokande}

Super-Kamiokande performed three runs between 1996 and 2007. The number of observed muons agrees well with the prediction from atmospheric neutrinos, there is no hint for an additional flux of neutrinos from the sun. We will now use the data of Super-Kamiokande I,~II and III to constrain dark matter annihilation in the sun.

Super-Kamiokande is able to measure the angle under which a muon is produced. At very high energies the muon inherits its direction of flight from the parent neutrino. Even for neutrinos in the GeV range the angle between neutrino and muon $\theta_{\mu\nu}$ is seldom larger than $30^\circ$. Therefore one can put especially strong constraints on the neutrino flux from the sun if one restricts the analysis to muons with $\theta_{\mu\odot}\leq 30^\circ$ where $\theta_{\mu\odot}$ is the angle between muon and sun-earth direction. (Note that $\theta_{\mu\odot}=\theta_{\mu\nu}$ for neutrinos coming from the sun).

For fully contained muon-like events there exist, unfortunately, only data on the total number of events. From figure 1 in~\cite{Wendell:2010md} we extract that 8596 events of this type have been observed during the time $t^{\text{FC}}=2806\:\text{d}$. The Monte Carlo prediction for atmospheric neutrino events was 8610.\footnote{The Monte Carlo prediction was calculated for the neutrino oscillation parameters set to the best fit point of the Super-Kamiokande analysis. A slight variation of $\theta_{13}$ would, however, only marginally affect the prediction as can be seen in the same figure.} This translates into a 90\% Poisson upper limit on the (dark matter induced) event rate
\begin{equation}\label{eq:maxeventsfc}
R^\text{FC}_\text{max} = 13.8 \yr^{-1}\;.
\end{equation}
For upward stopping muons the event distribution with respect to $\theta_{\mu\odot}$ can be taken from~\cite{Tanaka:2008zz,Tanaka:2009}. Considering only events with $\theta_{\mu\odot}\leq 30^\circ$, the observed number during a run time of $t^{\text{Stop}}=2828\:\text{d}$ was 53, the Monte Carlo atmospheric neutrino prediction 54 (see figure~1 of~\cite{Tanaka:2009}). The corresponding 90\% Poisson upper limit on the (dark matter induced) event rate is
\begin{equation}\label{eq:maxeventsstop}
R^{\text{Stop,}\,30^\circ}_\text{max} = 1.24 \yr^{-1}\;.
\end{equation}
The prediction for $R^{\text{Stop,}\,30^\circ}$ is given by~\eqref{eq:ratestopping} with the $\D\cos\theta_{\mu\nu}$ integration running from $\cos(30^\circ)$ to $1$ rather than from $-1$ to $1$.

\begin{figure}[t] 
  \begin{minipage}[b]{8.2 cm}
    \includegraphics[width=8.2cm]{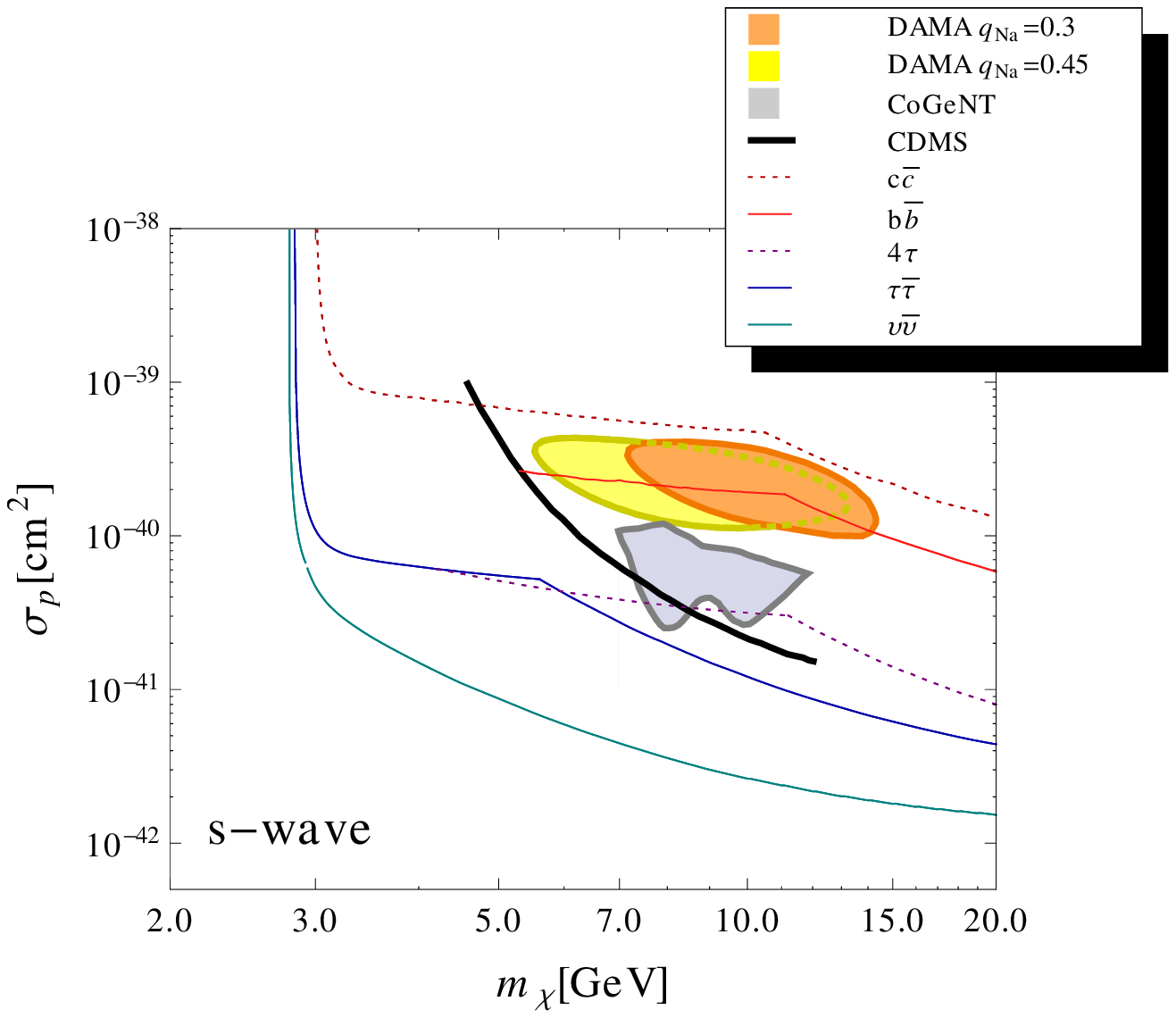}  
  \end{minipage}\hspace*{-0.8cm}
  \begin{minipage}[b]{8.2 cm}
    \includegraphics[width=8.2cm]{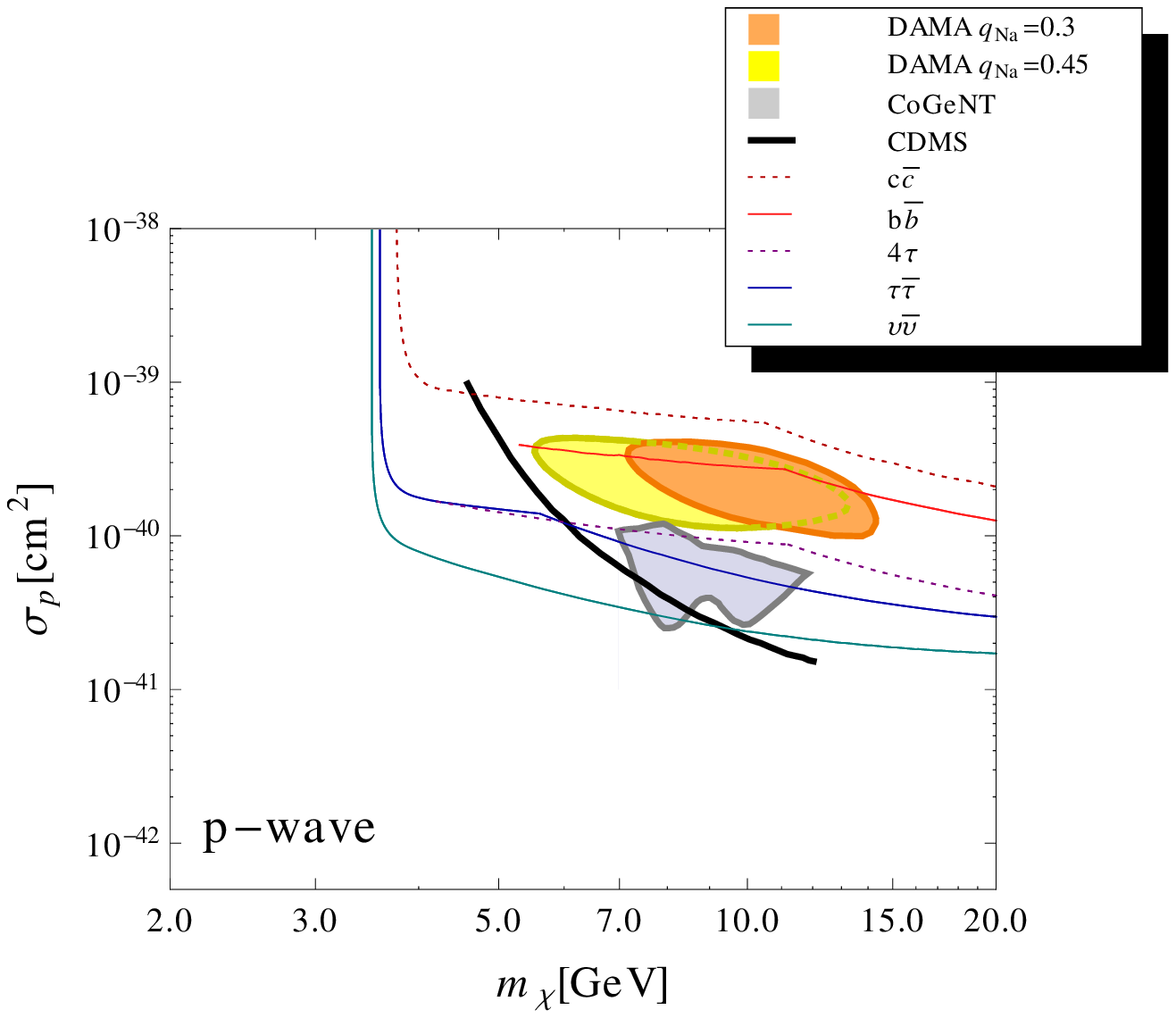}  
  \end{minipage}
  \caption{The Super-Kamiokande 90\% upper limits on the spin-independent WIMP nucleon cross section for various annihilation channels (equal couplings to proton and neutron are assumed). The annihilation cross section is chosen such that the thermal WIMP abundance matches the observed dark matter abundance for s-wave annihilation (left panel) and p-wave annihilation (right panel). At low WIMP mass the limits arise from fully contained events, at higher WIMP mass from upward stopping muons. Below the evaporation mass all constraints disappear rapidly. Also shown are confidence regions and limits from direct detection experiments.}
  \label{fig:limitssi}
\end{figure}

\begin{figure}[t] 
  \begin{minipage}[b]{8.2 cm}
    \includegraphics[width=8.2cm]{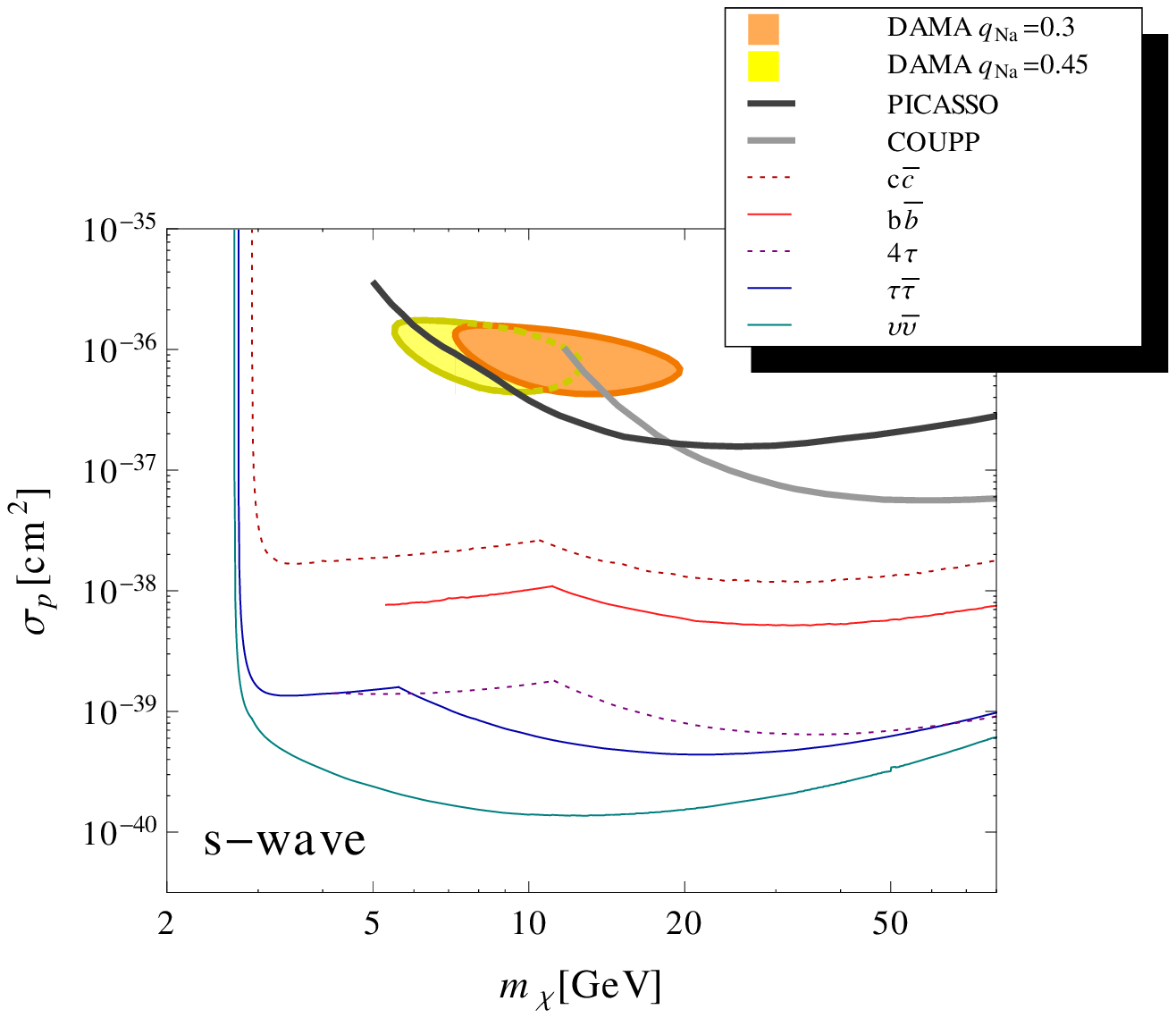}  
  \end{minipage}\hspace*{-0.8cm}
  \begin{minipage}[b]{8.2 cm}
    \includegraphics[width=8.2cm]{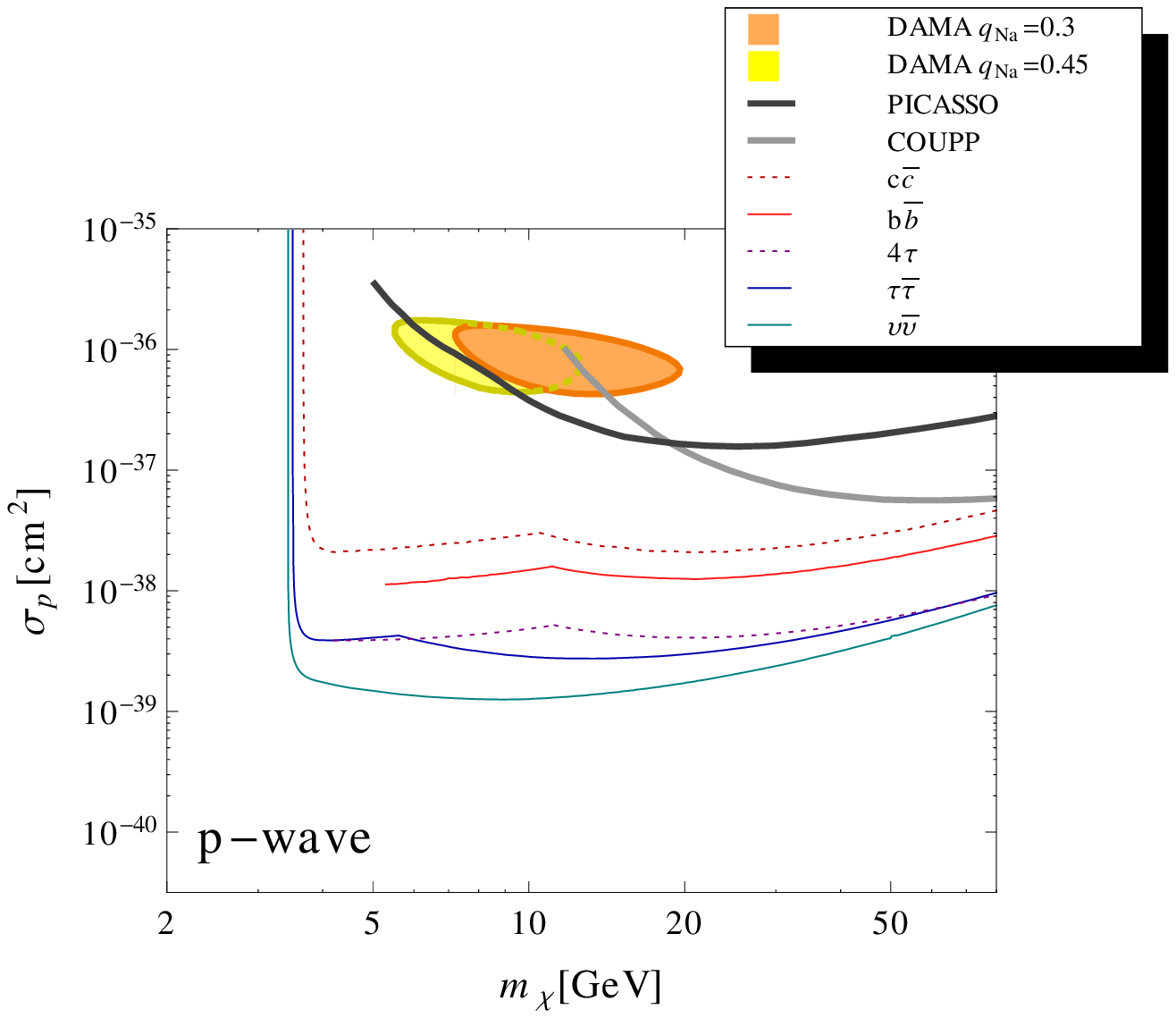}  
  \end{minipage}
  \caption{Same as figure~\ref{fig:limitssi} for spin-dependent scattering. Couplings only to protons are assumed.}
  \label{fig:limitssd}
\end{figure}

The limits~\eqref{eq:maxeventsfc} and~\eqref{eq:maxeventsstop} translate into limits on the spin-independent and spin-dependent WIMP nucleon cross section. The combined limits from fully contained events and upward stopping muons are shown together with confidence regions and constraints from direct detection experiments in figures~\ref{fig:limitssi} and~\ref{fig:limitssd}. We independently treat s-wave and p-wave annihilation (see section~\ref{sec:darkmattersun}).

In the spin-independent case we assumed equal couplings to protons and neutrons\footnote{A different weighting of couplings would merely lead to an overall shift of all limits and confidence intervals. This is due to the fact that all nuclei apart from hydrogen have roughly the same ratio of protons to neutrons. (And hydrogen does not contribute substantially to the capture in the sun for spin-independent scattering.)}, in the spin-dependent case couplings to protons only.\footnote{A non-zero coupling to neutrons would not affect the Super-Kamiokande limits, but the limits and confidence regions from direct detection experiments.} For DAMA we calculated the $3\sigma$ confidence region as described in~\cite{SchmidtHoberg:2009gn}\footnote{Spin-dependent scattering is not explicitly covered in~\cite{SchmidtHoberg:2009gn}, but the analysis is analogous. One only has to include the spin-structure functions which we take from~\cite{Ressell:1997kx}.} for two different values of the sodium quenching factor $q_{\text{Na}}$. The standard value is $q_{\text{Na}}=0.3$, but $q_{\text{Na}}=0.45$ is still within the range of experimental uncertainties and would allow to reconcile the DAMA and CoGeNT signals (see discussion in~\cite{Hooper:2010uy}). Channeling was assumed to be absent as suggested by~\cite{Bozorgnia:2010xy}. We used $600\km\s^{-1}$ for the galactic escape velocity as in the CoGeNT analysis. The CoGeNT favored region was taken from~\cite{Aalseth:2010vx}, the CDMS limit from~\cite{Ahmed:2010wy}. We did not try to fit the CRESST signal as the collaboration has not yet published their data. 

As can be seen the CDMS limit excludes the DAMA and most part of the CoGeNT favored region.\footnote{Important limits also arise from the XENON~10/100 experiments~\cite{Angle:2007uj,Aprile:2010um}. These depend, however, strongly on the low-energy behavior of the scintillation efficiency in liquid XENON which is subject to large uncertainties (see discussion in~\cite{Collar:2010gg}).} However, serious concerns about the CDMS analysis have been raised in~\cite{Collar:2011kf}. If we employ the limits from Super-Kamiokande we find that the CoGeNT and DAMA favored regions are excluded for WIMPs which annihilate dominantly into neutrino or tau pairs unless the annihilation cross section is velocity suppressed. Even for pure p-wave annihilation DAMA remains excluded. The quark channels are less constrained.
Our results can be used to restrict possible explanations to CoGeNT and DAMA further. Observations from the cosmic microwave background~\cite{Hutsi:2011vx} already disfavor thermal WIMPs with $m_\chi\lesssim 8\gev$ which annihilate into quarks, gluons, photons or electrons. This statement, however, only applies if the annihilation cross sections at freeze-out and CMB formation are equal. A possible loop-hole remains velocity suppressed annihilation.

Summing up the limits from indirect dark matter detection, the annihilation channels of light thermal WIMPs are tightly constrained. If we require a simultaneous explanation to CoGeNT and DAMA, dominant annihilation into muons remains an option. If the annihilation cross section is velocity suppressed all annihilation channels apart from the neutrino and tau channels remain viable. In~\cite{Kappl:2010qx} we discussed e.g. a model with p-wave annihilation of WIMPs into light Higgs pairs which subsequently decay into taus and charms. For this case the $4\tau$-limit alleviated by a factor corresponding to the branching fraction into taus $(\sim 1/2)$ applies, i.e. the model can explain the CoGeNT and DAMA signals without being in conflict with any indirect detection bounds. We also note that indirect detection limits may be modified by invoking non-standard astrophysics~\cite{Serpico:2010ae}.

Let us now turn to spin-dependent WIMP nucleus scattering which was introduced to reconcile DAMA with other direct detection experiments~\cite{Savage:2004fn}. The DAMA $3\sigma$ region is excluded by PICASSO~\cite{Archambault:2009sm} and COUPP~\cite{Behnke:2010xt} unless the quenching factor of sodium is slightly above the standard value $q_{\text{Na}}=0.3$. The Super-Kamiokande limits are substantially stronger than direct detection limits for all considered annihilation channels. Especially the DAMA region is ruled out for WIMPs which annihilate into neutrinos, taus, bottoms or charms, even if the branching fraction is only at the few percent level.

\section{Conclusion}

In this study we have analyzed WIMP annihilation in the sun and the corresponding neutrino signals. We have determined the expected event rate of upward stopping muons and fully contained muon-like events at Super-Kamiokande for various possible WIMP annihilation channels. This allowed us to derive strong constraints on the WIMP nucleon cross section especially for light WIMPs. We took into account the possibility of velocity suppressed annihilation and discussed separately the cases of spin-independent and spin-dependent WIMP nucleus scattering. Our results were then related to observations at direct dark matter detection experiments. 

For the spin-independent case we were able to exclude WIMPs which annihilate into tau or neutrino pairs as a possible source of the signals seen at DAMA and CoGeNT. This statement holds unless the annihilation cross section is velocity suppressed, but even then the DAMA region remains excluded for these channels. If one takes the existing limits from WMAP serious a simultaneous solution to DAMA and CoGeNT with a thermal WIMP requires either annihilation dominantly into muons\footnote{Annihilation into particles outside the Standard Model can also not be excluded.} or a velocity suppressed annihilation cross section.

For the spin-dependent case we found that the Super-Kamiokande limits are substantially stronger than those from direct detection in the considered mass range ($m_\chi \leq 80\gev$). The DAMA favored region is ruled out for annihilation into neutrinos, taus, bottoms and charms. The constraints -- even for velocity suppressed annihilation -- are so strong that a solution to DAMA with a (spin-dependent) thermal WIMP seems rather difficult.

\section*{Acknowledgments}
We would like to thank Carolin B. Br\"auninger, Michael Ratz and David Tran for helpful discussion as well as Jes\'us Torrado Cacho for his assistance in handling the IDL Astronomy User's Library.
This research was supported by the DFG
cluster of excellence Origin and Structure of the Universe, and the
SFB-Transregio 27 "Neutrinos and Beyond" by Deutsche
Forschungsgemeinschaft (DFG).

\bibliography{neubib}
\bibliographystyle{OurBibTeX}
\end{document}